\documentclass[manuscript]{aastex}
\usepackage{gensymb}

\slugcomment{Oct 3rd, 2010}

\shorttitle{DISCS:  II. Southern Sky Protoplanetary Disk Data}
\shortauthors{\"Oberg et al.}

\begin{document}

\title{Disk Imaging Survey of Chemistry with SMA: II. 
Southern Sky Protoplanetary Disk Data and Full Sample Statistics}

\author{
Karin I. \"Oberg\altaffilmark{1}, 
Chunhua Qi}
\affil{Harvard-Smithsonian Center for Astrophysics, 60 Garden Street, Cambridge, MA 02138, USA} 

\and

\author{
Jeffrey K.J. Fogel, 
Edwin A. Bergin}
\affil{Department of Astronomy, University of Michigan, Ann Arbor, MI 48109, USA}

\and

\author{
Sean M. Andrews, 
Catherine Espaillat\altaffilmark{2}, 
David J. Wilner}
\affil{Harvard-Smithsonian Center for Astrophysics, 60 Garden Street, Cambridge, MA 02138, USA}

\and

\author{Ilaria Pascucci}
\affil{Department of Physics and Astronomy, Johns Hopkins University, 3400 N. Charles Street, Baltimore, MD 21218, USA}

\and

\author{Joel H. Kastner}
\affil{Center for Imaging Science, Rochester Institute of Technology, 54 Lomb Memorial Drive, Rochester, NY 14623, USA}

\altaffiltext{1}{Hubble Fellow}
\altaffiltext{2}{NSF Astronomy \& Astrophysics
Postdoctoral Fellow}

\begin{abstract}
This is the second in a series of papers based on data from DISCS, a
Submillimeter Array observing program aimed at spatially and spectrally
resolving the chemical composition of 12 protoplanetary disks.
We present data on six Southern sky sources -- IM Lup,
SAO 206462 (HD 135344b), HD 142527, AS 209, AS 205 and V4046 Sgr -- which complement
the six sources in the Taurus star forming region reported previously.
CO 2--1 and HCO$^+$ 3--2 emission are detected and resolved in all disks
and show velocity patterns consistent with Keplerian rotation.
Where detected, the emission from DCO$^+$ 3--2, N$_2$H$^+$ 3--2,
H$_2$CO 3$_{0 \: 3}-2_{0 \: 2}$ and 4$_{1 \: 4}-3_{1 \: 3}$, HCN 3--2
and CN 2$_{3 \: 3/4/2}-1_{2 \: 2/3/1}$ are also generally spatially resolved.
The detection rates are highest toward the M and K stars, while the F star
SAO 206462 has only weak CN and HCN emission, and H$_2$CO alone is detected toward HD 142527. These findings together with the statistics from the previous Taurus disks, support the hypothesis that high detection rates of many small molecules depend on the presence of a cold and protected disk midplane, which is less common around F and A stars compared to M and K stars.
Disk-averaged variations in the proposed radiation tracer CN/HCN are found
to be small, despite two orders of magnitude range of spectral types and accretion rates. In contrast, the resolved images suggest that the CN/HCN emission ratio
varies with disk radius in at least two of the systems. There are no clear observational differences
in the disk chemistry between the classical/full T Tauri disks and transitional
disks. Furthermore, the observed line emission does not depend on measured accretion
luminosities or the number of infrared lines detected, which suggests that
the chemistry outside of 100~AU is not coupled to the physical processes that drive the chemistry in the innermost few AU.

\end{abstract}

\keywords{protoplanetary disks; astrochemistry; stars: formation; ISM: molecules; techniques: high angular resolution; radio lines: ISM}

\section{Introduction}

\noindent Molecular abundances in protoplanetary disks provide important clues to the chemical evolution and the physical conditions prevalent during star- and planet-formation. The chemical composition of disks will affect the matter incorporated into planetesimals and is therefore of great pre-biotic interest -- for example, observations of comets \citep{Crovisier04} show that astrochemical pathways to molecular complexity exist. 

Disk chemistry is an active topic, but observations are difficult because of arc-second angular sizes and intrinsically low gas-phase molecular abundances. Prior to the Disk Imaging Survey of Chemistry with the SMA (DISCS), molecules other than CO and HCO$^+$ were observed in the millimeter in single-dish studies of seven different disks, including 11 different species detected toward DM Tau \citep{Kastner97,Dutrey97,vanDishoeck03,Thi04,Kastner08,Fuente10}. In addition TW Hya has been mapped in DCO$^+$, HCN and DCN, and DM Tau and LkCa 15 in N$_2$H$^+$, C$_2$H and H$_2$CO \citep{Qi03,Aikawa03,Dutrey04,Dutrey07,Qi08,Henning10}. This limited data set has provided insight into disk chemistry, but a larger sample of molecular distributions in disks is needed to explore chemical trends with respect to stellar properties and disk evolution.

The aim of  DISCS is to constrain the impact of the disk structure and radiation environment on the disk chemistry through a homogenous survey of spatially and spectrally resolved molecular emission in twelve protoplanetary disks. The star+disk systems have been chosen to span a range of stellar spectral types, disk accretion rates and disk density profiles, including both classical or full disks and disks with gaps and holes \citep[pre-transitional and transitional disks: ][]{Strom89,Espaillat07}. The targeted molecules are the simple species that previous studies suggested may be detectable (CO, HCO$^+$, DCO$^+$, CN, HCN, DCN, N$_2$H$^+$, C$_3$H$_2$, H$_2$CO and CH$_3$OH). The first part of the survey was presented by \citet[][from now on Paper I]{Oberg10c} with 3--8 different molecules detected in six disks in the Taurus molecular clouds. 

The second part of the survey contains six protoplanetary disks in the Southern sky: IM Lup, AS 205, AS 209, V4046 Sgr, SAO 206462 and HD 142527.  The disk sample is presented in $\S$\ref{sec:sample} with special attention to the properties that may affect the disk chemistry. The spectral set-ups and observations are described in $\S$\ref{sec:obs}. The extracted spectra and position-velocity diagrams of all lines are shown in $\S$\ref{sec:res}, together with moment maps of CO, HCO$^+$, CN and HCN. The new Southern disk data is combined with the previously published Taurus disk data to compare the chemistry in full and
transitional disks, and to investigate CN/HCN variations across the sample and within individual disks. The detection rates in this complete DISCS sample as well as source-to-source differences are discussed in $\S$\ref{sec:disc}, especially with reference to previous studies, the radiation chemistry, disk structure effects, and inner vs. outer disk chemistry.

\section{The Disk Sample\label{sec:sample}}

\noindent The complete 12-source DISCS sample was chosen to assess the impact of spectral type or stellar irradiation field, accretion luminosity and structure on the disk chemistry. It consists of six disks in Taurus (Paper I) and six disks in the Southern sky (Table \ref{tbl:star}). In this paper, the six new disks are referred to as `the Southern sample' and the complete 12-source sample of disks as `the DISCS sample'. The DISCS sample is biased toward large disks, since disks smaller 
than a few hundred AU are not spatially resolved by the SMA in the compact 
configuration. The observations resolve molecular distributions at radii beyond 100-400~AU, depending on the source distance and beam shape, which means that there is a large gap in radii between these millimeter observations and infrared disk chemistry observations that typically probe the inner disk atmosphere out to a few AU. The sources were selected from disks previously mapped in CO \citep{Ohashi08, Andrews09, Panic09, Rodriguez10}, and thus may be biased toward gas-rich disks. All disks are clearly isolated from the parent cloud emission, which reduces confusion, but may result in an evolutionary bias. In Paper I, several of the disks were known hosts of  organic molecules (DM Tau, LkCa 15 and MWC 480), while the Southern sample contains disks of unknown chemical complexity, except for V4046 Sgr, which is known to exhibit strong HCO$^+$, CN, and HCN emission \citep{Kastner08}.

Stellar luminosity, accretion luminosity, X-rays, the interstellar irradiation field, disk geometry, disk gaps and holes, dust settling and growth may all affect the chemistry in the disk. Their predicted impacts are discussed in detail in Paper I. The complete DISCS survey encompasses two M stars, six K stars, three F stars and one A star, with reported stellar luminosities between 0.3 and 69 L$_{\odot}$. This range allows us to investigate whether stellar luminosity is an important factor for the disk chemistry, even if several factors contribute to the chemical evolution. A second source of radiation, mass accretion, also varies considerably within the DISCS sample. Among the K stars, the mass accretion rates are $3-90\times10^{-9}$ M$_{\odot}$ yr$^{-1}$, with AS 205 and AS 209 at the higher end and IM Lup and the transition disks in Paper I at the lower end. X-ray luminosities are generally higher toward T Tauri stars compared to Herbig Ae/Be stars. Published X-ray luminosities exist for some of the disks, and the possible impact of X-rays on the observed chemistry will be discussed in upcoming papers on the best constrained systems. 

The most obvious variation within the DISCS sample is the difference between full disks and transition disks, where planetary systems may be the source of large gaps and holes \citep[e.g.][]{Lubow06}. The complete survey is equally divided between (pre-)transitional disks with gaps or holes (DM Tau, LkCa 15, GM Aur, V4046 Sgr, HD 142527, SAO 206462) and full disks (AA Tau, IM Lup, AS 205, AS 209, CQ Tau, MWC 480). (Pre-)transitional disks are generally identified through SED modeling \citep{Espaillat07}, but several disks have large enough holes to be imaged. Of the Southern (pre-)transitional disks,  V4046 Sgr has an estimated small 0.2~AU hole \citep{Jensen97}, SAO 206462 has a gap between $\sim$0.3 and $\sim$50~AU \citep{Grady09} and HD 142527 has a hole with radius 100-150~AU \citep{Fukagawa06}.

\section{Observations\label{sec:obs}}

\subsection{Spectral Setups}

\noindent Two frequency setups per source were selected to cover 4 to 8 spectral lines 
in each setup, at the same time providing continuum observations.
The targeted molecules are DCO$^+$ and DCN (probing deuterium chemistry and cold gas), 
CN and HCN (probing photochemistry), HCO$^+$ (probing ionization), N$_2$H$^+$ (potential CO freeze-out tracer), H$_2$CO 
and CH$_3$OH (potential grain chemistry products), $c$-C$_3$H$_2$ (carbon-chain chemistry) and CO (the disk kinematic tracer). 
Tables ~\ref{tbl:setup1} and ~\ref{tbl:setup2} summarize the two spectral 
setups, centered at 1.1 mm and 1.4 mm, respectively. 

The SMA correlator covers $\sim$4~GHz bandwidth per sideband 
using two Intermediate Frequency (IF) bands with widths of 1.968 GHz.
The first IF band is centered at 5 GHz, and the second IF band is centered 
at 7 GHz from the Local Oscillator (LO). Each band is divided into 24 slightly overlapping ``chunks'' of 
104 MHz width, which can have different spectral resolution.  For each spectral setting, the correlator was configured to increase the spectral resolution on the key species with 128--512 channels per chunk. Chunks containing weaker lines were binned for higher sensitivity, while still recovering sufficient kinetic information. The remaining chunks were covered by 64 channels each and used to measure  the continuum. The continuum visibilities in each sideband and each IF band  were generated by averaging the central 82 MHz in all line-free chunks.

\subsection{Data Acquisition and Calibration}

The DISCS sources were observed from 2010 April through August with the 
compact or compact-north configuration of the Submillimeter Array (SMA) interferometer 
(Ho et al. 2004) at Mauna Kea, Hawaii. For each observation, at least six 
of the eight 6~m SMA antennas were available, spanning baselines of 16--77(139)~m 
(Table ~\ref{tbl:obs}).  V4046 Sgr was partially observed with one IF band in an earlier program February--September 2009 (PI J. Kastner) and therefore contains a different H$_2$CO 3--2 transition. The CO data acquired within this program was presented in \citet{Rodriguez10}. 
The observing sequence interleaved disk targets and two quasars in an 
alternating pattern. A group of three quasars was used 
for gain calibration -- J1604-446, J1626-298 and J1733-130 -- depending on their fluxes at the time of the observations and proximity to the disk. The observing conditions 
were generally very good, 
with $\tau_{\rm 225{\:GHz}}\sim0.05-0.1$ and stable atmospheric phase.

The data were edited and calibrated with the IDL-based MIR software 
package\footnote{http://www.cfa.harvard.edu/$\sim$cqi/mircook.html}.
The bandpass response was calibrated with observations of Uranus and the bright quasars available (3C 454.3 and 3C 273). Observations of Uranus, Neptune and Callisto provided the absolute scale for the calibration of 
flux densities. The typical systematic uncertainty in the absolute flux 
scale is $\sim$10\%. Continuum and spectral line images were generated 
and CLEANed using MIRIAD. 

\section{Results}\label{sec:res}


\noindent This section presents new data from the six Southern disks, but the results from Paper I on six Taurus disks are used extensively to investigate general trends in disk chemistry with disk structure and radiation fields. The disk chemical contents are presented and discussed in terms of line fluxes, which depend on both molecular column densities and excitation conditions. For lines that are optically thin, not CO and HCO$^+$, and originate in similar regions in all disks, the line flux ratios should be proportional to the molecular abundance ratios. T Tauri and Herbig Ae stars may not have comparable emission regions, \citep{Thi04}, however, and full radiative-transfer modeling is required to confirm observed chemical trends that are based on flux ratios.

\begin{figure*}[htp]
\epsscale{1.0}
\plotone{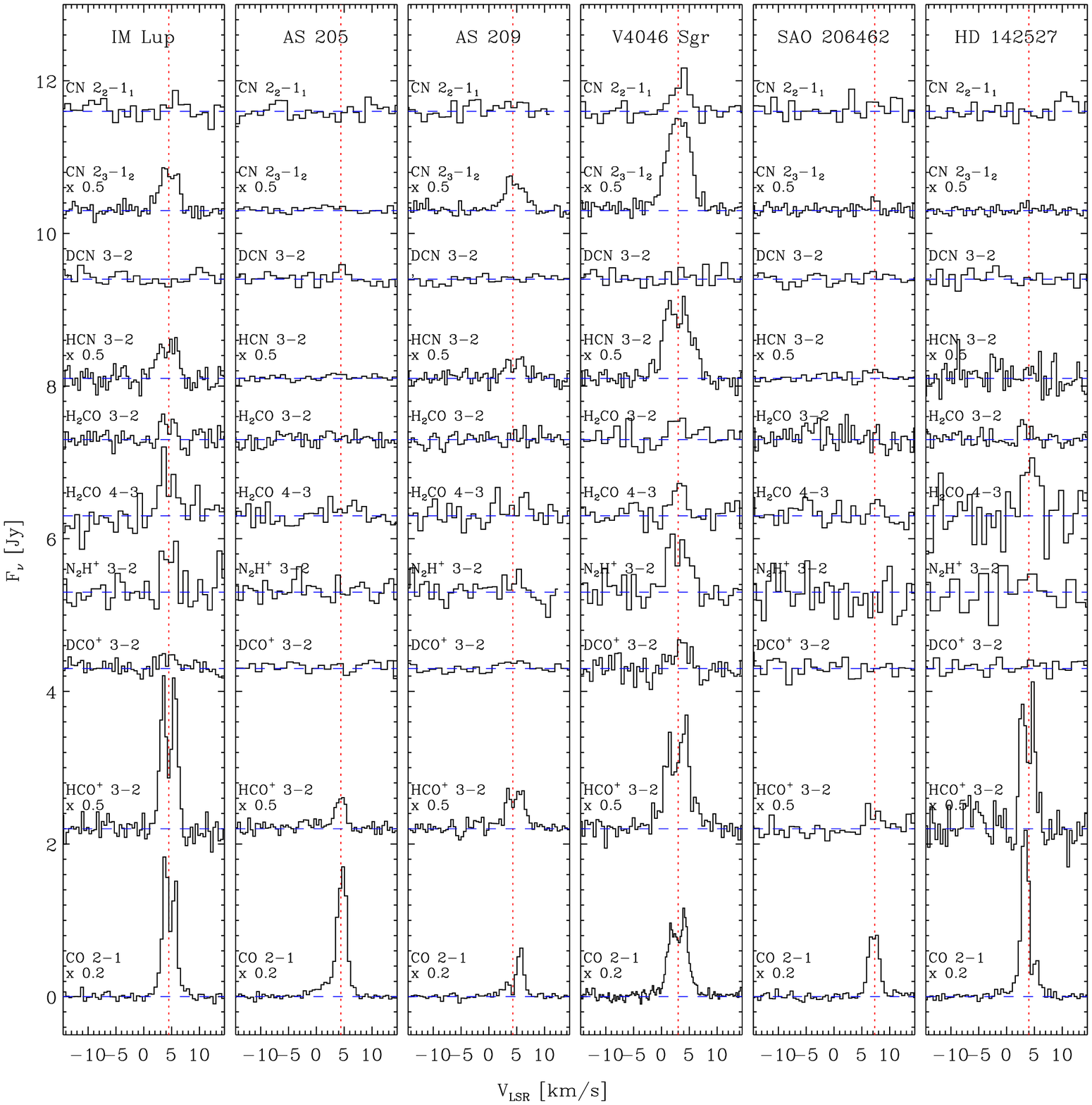}
\caption{Spatially integrated spectra displaying the chemical variation between the different disks. The line profiles of CO, HCO$^+$, HCN and one of the CN lines have been scaled for visibility. \label{fig:spec}}
\end{figure*}

Figure \ref{fig:spec} shows the integrated spectra of the targeted molecules toward IM Lup, AS 205, AS 209, V4046 Sgr, SAO 206462 and HD 142527. The spectra are also tabulated in Table \ref{tbl:spec}. The spectral lines are extracted from emission maps using elliptical masks produced by fitting a Gaussian profile to the CO integrated intensity maps 
toward each source. The size of the mask is scaled down for a few lines, to optimize the signal-to-noise without losing any significant emission, as in Paper I.

The line profiles are generally consistent with rotating disks, but most of the CO 2--1 lines are asymmetric, and the AS~205 line seems to have a wing. The peak asymmetries are probably dominated by cloud contaminations, since the CO 2--1 profile is more severely affected compared to other lines. The AS 205 line profile may be additionally affected by the existence of a companion \citep{Andrews09}. Dust continuum images of HD 142527 show some evidence for non-axisymmetric
structure \citep{Fujiwara06}, which may explain the asymmetries in all spectral
line emission detected toward this system. It is possible that winds or unknown outflows could contribute emission at high velocities in some systems, and detailed modeling will be needed
to test if such additional components are required to explain the data.

The integrated line fluxes toward the southern sources are listed in Table \ref{tbl:int}. The 2-$\sigma$ upper limits are calculated from the rms in the disk-integrated spectra and the FWHM of the CO 2-1 line according to the standard formula: $\sigma=rms\times FWHM / \sqrt{n_{\rm ch}}$, where $n_{\rm ch}$ is the number of channels across the CO 2-1 feature at FWHM. Using the CO 2-1 FWHM is supported by the similarity in the line widths for different transitions toward the same disk in Fig. \ref{fig:spec}. Because of variable observing conditions and integration times, the integrated flux upper limits vary between 0.2 and  1.0 Jy km s$^{-1}$. The typical rms is almost a factor of two larger compared to Paper I because of the low declination of the Southern sources,  resulting in shorter integration times and larger atmospheric columns. 

CO and HCO$^+$ alone are detected toward all Southern sources, CN and HCN toward four disks each,  and DCO$^+$, N$_2$H$^+$ and H$_2$CO toward IM Lup, AS 209 and V4046 Sgr. H$_2$CO is also observed toward HD 142527. DCN, CH$_3$OH and $c$-C$_3$H$_2$ are not detected toward any of the sources in this sample. Detections are here defined as $>$3$\sigma$ integrated line intensities and 2$\sigma$ contours present in at least three channel maps. Tentative detections have integrated line intensities of 2--3$\sigma$ and 2$\sigma$ contours in at least one channel map.  Among the detections, HCO$^+$, CN and HCN line intensities span more than an order of magnitude. The CO 2--1 flux varies less because of its high optical depth toward most disks, and the minor species DCO$^+$, N$_2$H$^+$ and H$_2$CO vary by only factors 3--4 because of low S/N.

Within the complete DISCS sample, only AS 205 and HD 142527 lack both CN and HCN emission. H$_2$CO emission is weaker than the CN and HCN lines in all sources where all three species are detected. It is therefore curious that the two H$_2$CO lines are detected toward  HD 142527, and this may be connected with its irregular disk structure.  AS 205 has a companion, which is unresolved here, which has resulted in a truncated disk \citep{Andrews09}. This may explain the lack of line detections, despite the strong CO 2-1 emission.

In terms of line flux, most upper limits are lower than typical detected fluxes. The upper limits on the DCO$^+$, N$_2$H$^+$ and H$_2$CO fluxes toward AS 205, AS 209, SAO 206462 and HD 142527 are all below the Southern source sample medians of 0.41, 1.59, 1.37 and 0.53 Jy km s$^{-1}$, respectively. CN and HCN flux upper limits are generally an order order of magnitude below the sample mean. This implies that most non-detections are significant and due to either chemical differences between the disks, or to differences in the sizes of the emission regions. The latter probably explains the paucity of line detections toward AS 205 and SAO 206462, based on their small CO emission regions (see below). In the Taurus sample a small emission region also explains the low line fluxes toward AA Tau and CQ Tau. The Herbig Ae stars HD 142527 and MWC 480 have large disks, however, and there seem to be real chemical differences between these disks and the large T Tauri disks.

\subsection{Spatial distributions of molecules within the Southern disks}

\begin{figure*}
\epsscale{1.0}
\plotone{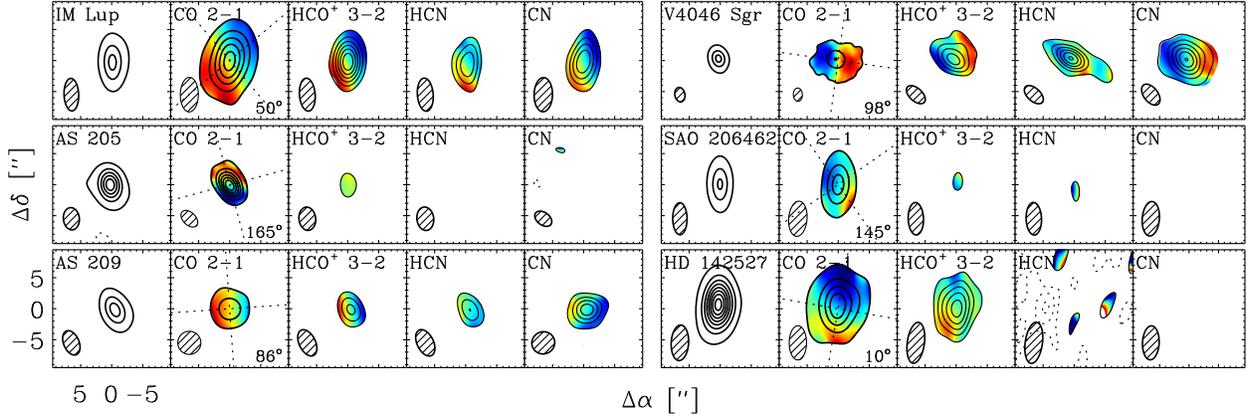}
\caption{267 GHz continuum and zero- and first moment maps of CO 2-1, HCO$^+$ 3-2, HCN 3-2 and CN 2-1. The continuum contours are 0.03 Jy followed by 0.1 Jy steps. In the moment maps the contours are 1.0 (then steps of 2.5), 1.0, 0.6 and 0.6 Jy km s$^{-1}$  beam$^{-1}$ for CO, HCO$^+$, HCN and CN respectively, except for V4046 Sgr, where the contour steps are 50\% larger. The velocity gradient is defined over the same range for all molecules toward each source. The position angles are overplotted and also listed in the CO 2--1 panels (90$\degree$ discrepancies with Table \ref{tbl:star} are due to orthogonal P.A. definitions). \label{fig:mom}}
\end{figure*}

\noindent Figure \ref{fig:mom} shows dust maps and integrated intensity and first moment maps for CO, HCO$^+$, HCN and CN for each disk in the Southern source sample. The dust continuum flux densities at $\sim$1.1~mm and 1.4~mm vary by factors of 
4--17 among these sources, with HD 142527 the strongest and SAO 206462 the weakest. As in the Taurus sample, there is not a one-to-one  
correspondence between the strength of the dust continuum emission
and the CO 2-1 line emission, nor between CO 2--1 and the emission from the other three major lines. The lack of a correlation between CO 2--1 and dust emission does not imply a lack of correlation between gas and dust masses however, since the dust emission is optically thin while the CO emission is optically thick, and the CO flux may also be further reduced because of CO freeze-out at low temperatures.

The first-moment maps show the rotation pattern, and where HCO$^+$, CN and HCN emission are resolved (IM Lup, AS 209 and V4046 Sgr) there seems to be more HCO$^+$ present at high velocities, and thus smaller radii compared to CN. Quantifying this difference requires radiative transfer because the CN emission is a line triplet. Table \ref{tbl:size} instead lists the FWHM of each of the line emission regions. The FWHM are derived from Gaussian fits in the UV-plane using MIRIAD, where the visibilities have been averaged over the velocity range where emission is detected. Since the emission regions differ between different velocity channels this is a crude size estimate. It is however adequate to investigate relative emission region sizes for different lines toward each disk.  The CO 2--1 Gaussian fits are also used together with literature source distances to estimate the CO emission radii listed in Table \ref{tbl:det}. Based on the Gaussian fits, the CN emission regions are significantly larger than the HCO$^+$ regions toward DM~Tau, IM Lup, AA~Tau and LkCa 15. This is consistent with models by \citet{Aikawa02}, which show a steep decline in HCO$^+$ with radius because of the onset of CO freeze-out, while CN abundances are constant or increase with radius.

\begin{figure*}[htp]
\epsscale{1.0}
\plotone{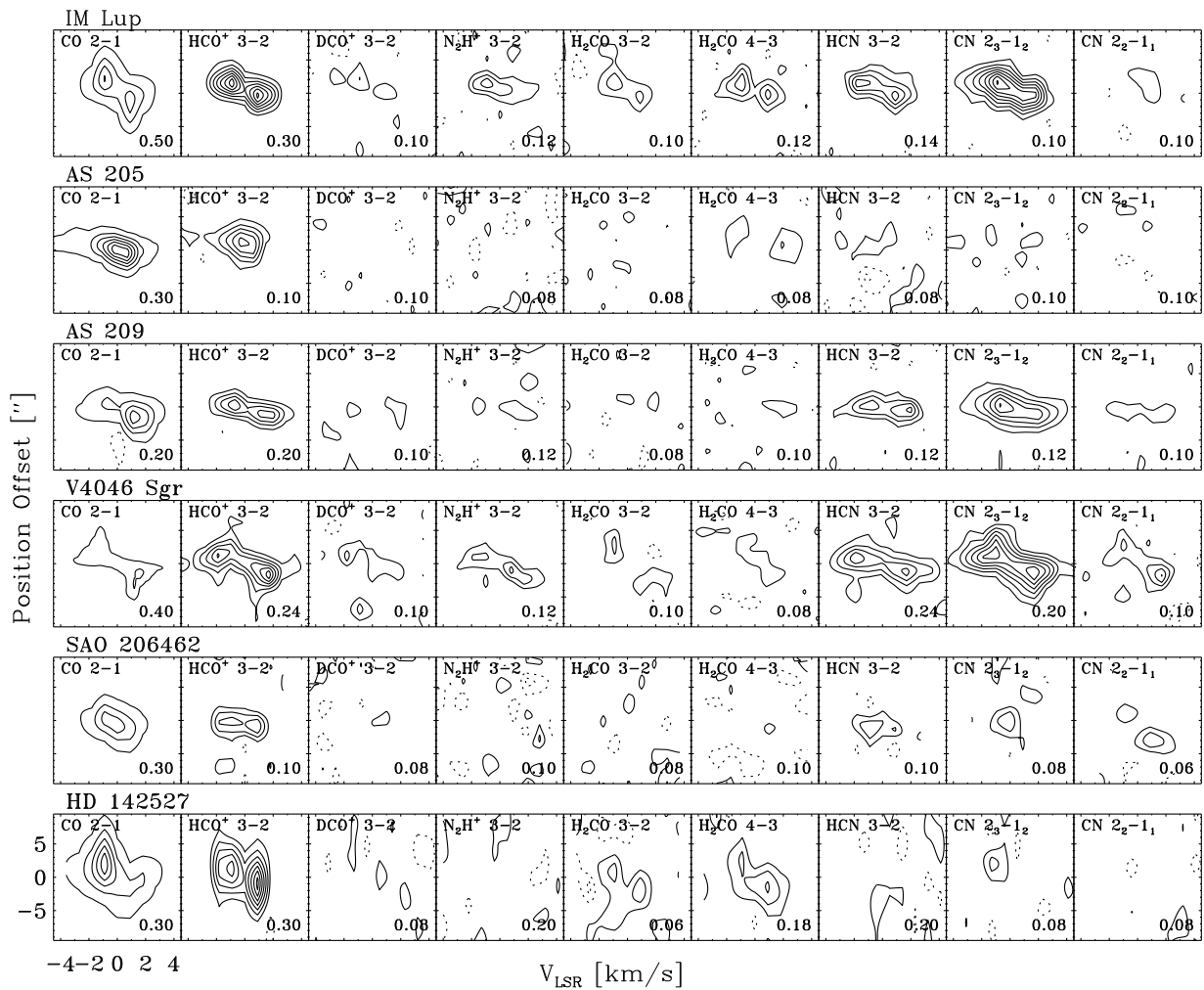}
\caption{Position-velocity diagrams for all lines detected toward at least one source. The contour levels are listed in the bottom right corner of each panel, except for CO where the steps above the first contour are three times larger for visibility. \label{fig:pv}}
\end{figure*}

Position-velocity (P-V) diagrams provide a more compact visualization of the distribution of molecules with radius; the diagrams show the line flux as a function of velocity offset and position along the disk major axis (Fig. \ref{fig:pv}). The CO major axes position angles are taken from the literature when available (IM Lup, AS 205, AS 209 and V4046 Sgr values listed in Table \ref{tbl:star}),  but for SAO 206462 and HD 142527 the axes are directly derived from the rotation pattern in the moment maps to 145 and 10$\degree$, respectively. The 145$\degree$ for SAO 206462 agrees with \citet{Grady09}, who found a P.A. for the dust disk of 55$\pm$5$\degree$ from NICMOS coronagraphic imaging and \citet{Pontoppidan08b} who found 56$\pm2\degree$ from spectroastrometric imaging of the gas close to the star. For completeness, Appendix A contains P-V diagrams for the Taurus disks from Paper I.

Line emission  from CO, HCO$^+$, CN and HCN, and the detected minor species are resolved and  consistent with disk rotation toward IM Lup, AS 209, V4046 Sgr and HD 142527, while the emission from SAO 206462 and AS 205 seem unresolved in all but CO. The PV-diagrams toward V4046 Sgr of the cold chemistry tracers DCO$^+$, N$_2$H$^+$ and H$_2$CO are noteworthy, since they reveal different distributions: the H$_2$CO emission seems to peak at a larger {\sl position} offset compared to the other molecules, and the DCO$^+$ emission seems to peak at larger {\sl velocity} offset, corresponding to a smaller emission radius, than N$_2$H$^+$.  There is not enough S/N to directly resolve whether these differences are significant, but this is suggestive that the three species trace different types of cold chemistry regions. N$_2$H$^+$ and possibly H$_2$CO require CO freeze-out to become abundant and their emission is expected to be related to cold grains. In contrast, DCO$^+$ forms in the gas-phase at low temperatures and its abundance should only depend on the gas temperatures. Gas and grain temperatures may be thermally decoupled in low-density regions \citep[e.g.][]{Aikawa06}, such as disk atmospheres. In disks with tapered density structures \citep{Hughes08,Andrews09}, the outer disk may also have low enough densities for grain and gas temperatures to decouple, resulting in gas that is heated by the Interstellar Radiation Field (ISRF) to several factors above the grain temperatures. A lack of cold gas, but not of cold grains, in the outer disk would explain why DCO$^+$ seems to have a smaller emission radius compared to N$_2$H$^+$ and H$_2$CO. Confirming different distributions of the cold chemistry tracers requires radiative transfer modeling, but these P-V diagrams already show the potential of constraining the chemistry and disk density profile through comparison between the spatial distributions of different emission lines.

\subsection{CN and HCN in the complete DISCS sample}

\begin{figure}[htp]
\epsscale{0.7}
\plotone{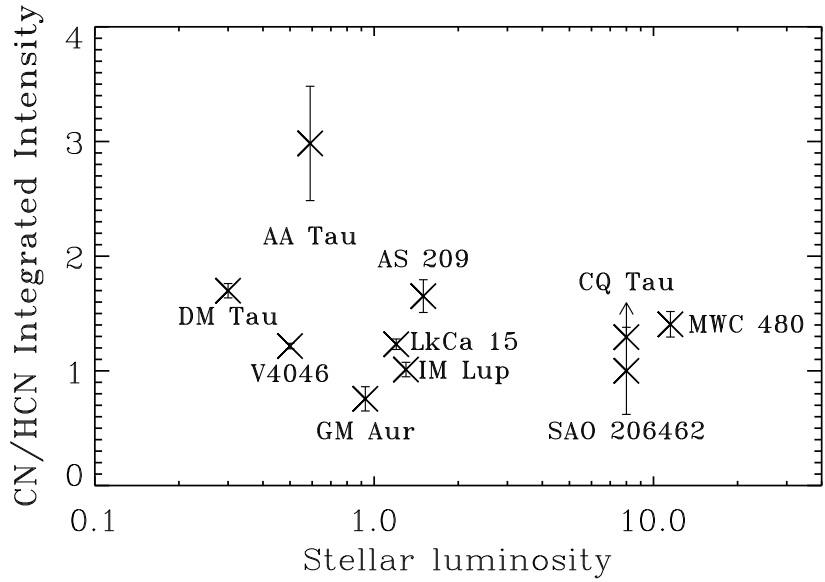}
\caption{The CN/HCN integrated flux ratio in the DISCS sample versus stellar luminosity. \label{fig:cn_hcn}}
\end{figure}

\noindent The CN/HCN abundance ratio is a proposed tracer of UV field whether dominated by stellar or accretion luminosity. However, Figure \ref{fig:cn_hcn} shows that the CN/HCN integrated flux ratio is almost constant across the complete DISCS sample at 1.4$\pm$0.6, with AA Tau as the sole outlier with a ratio of $\sim$3. This is consistent with the results from a smaller disk sample in \citet{Kastner08}. The flux is not directly proportional to the abundances because the HCN line may be modestly optically thick -- the strong CN 2--1 line is nearly optically thin from comparison with upper limits on the $\sim$10 times weaker CN 2--1 singlet line (see Paper I for a more detailed discussion) -- and because the emission regions for the two molecules may not be identical toward different disks. Still Fig. \ref{fig:cn_hcn} suggests that the disk averaged CN/HCN ratio is not a good tracer of the disk radiation environment. In addition to the emission ratios, there is one ratio limit toward CQ Tau, where only CN is detected. This limit is consistent with the detections.

\begin{figure}[htp]
\epsscale{0.4}
\plotone{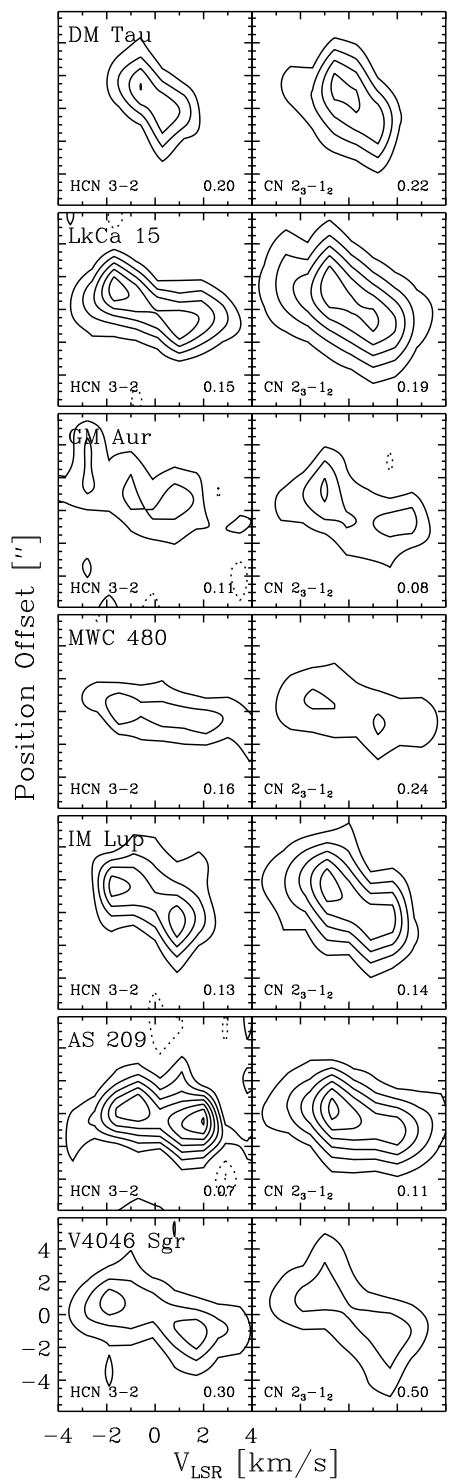}
\caption{P-V diagrams of the CN and HCN emission toward disks where both are resolved. Contours are in steps of 12.5--20\% of the peak flux. \label{fig:pv_cn}}
\end{figure}

The spatially resolved data does suggest that the CN/HCN ratio varies across some of the disks (Fig. \ref{fig:pv_cn}). The Gaussian fits to CN and HCN line emission from these disks show that toward two of the disks, V4046 Sgr and LkCa 15, the semi-major axes are significantly larger for CN compared to HCN; the differences in FWHM exceed the combined uncertainties (Table \ref{tbl:size}). For the other five sources the fits are inconclusive because of insufficient resolution or sensitivity. There is an apparent further extent of CN in velocity space compared to HCN, which normally implies more material present at smaller radii. The strong CN line is however a triplet, which results in confusion between contributions from material at different velocities and emission at slightly different fundamental frequencies.

\subsection{Full disks versus transitional disks}

The complete DISCS sample contains six T Tauri stars with large $>$400~AU disks, which are all detected in most lines. Two of these disks have no resolved holes (IM~Lup and AS~209), one has a sub-AU hole (V4046~Sgr) and three have 3-60~AU holes or gaps (DM Tau, GM Aur and LkCa 15) \citep{Calvet05,Espaillat10}. This sub-sample is used to investigate whether there are any systematic differences between full and (pre-)transitional disks. 

\begin{figure}[htp]
\epsscale{1.0}
\plotone{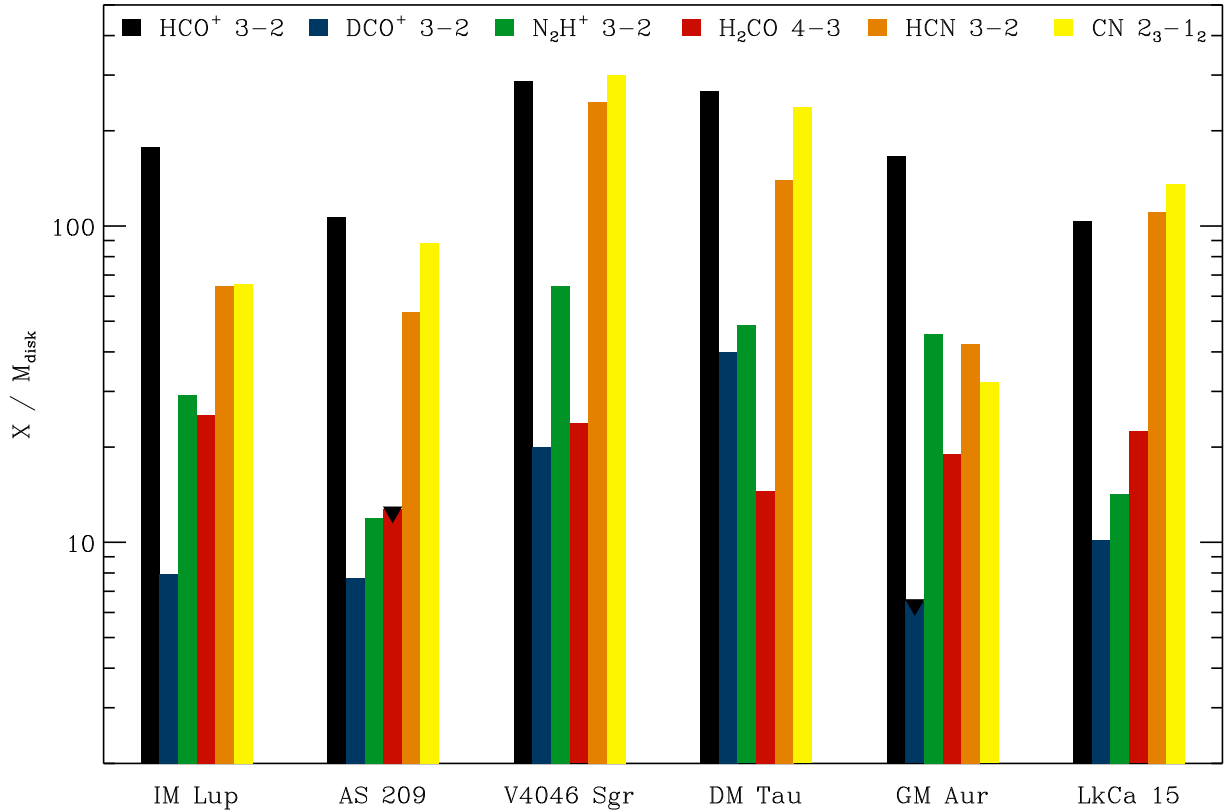}
\caption{Integrated intensities of six lines normalized to the disk masses and the source distances toward IM Lup, AS 209, V4046 Sgr, DM Tau, GM Aur and LkCa 15,  illustrating the lack of a systematic differences in the molecular emission pattern toward full and (pre-)transitional disks. The sources are ordered from full disks to disks with larger and larger holes and gaps. \label{fig:trans}}
\end{figure}

Figure \ref{fig:trans} shows the integrated emission of six of the detected lines (excluding one H$_2$CO, one CN and the DCN line) normalized to the disk masses and the Taurus distance. Except for the weak CN, HCN and DCO$^+$ emission toward GM Aur, the normalized line intensities are remarkably similar toward these very different disks. In particular there is no  pattern in the normalized line fluxes when progressing from the full disks to the disks with the biggest gaps and holes. There is thus no evidence from the disk averaged molecular emission that the outer disk chemistry is affected by disk clearing.

In addition to the variable disk structure, the mass accretion rates vary by two orders of magnitude among the sources with DM Tau and IM Lup  on the low end and AS 209 on the high \citep{Bergin04, Pinte08, Andrews09}. The accompanying accretion luminosity can be calculated from $L_{\rm acc} = G M_{\rm*}\dot{M}/R_{\rm*}$ \citep{Calvet04} and since the T Tauri stars have comparable stellar masses $M_{\rm*}$ and radii $R_{\rm*}$ the two order of magnitude differences in accretion rates result in two orders of magnitude differences in accretion luminosities. Figure \ref{fig:trans} reveals no clear difference between the low- and high-rate accretors, suggesting that the outer disk chemistry is not regulated by accretion luminosity. It is important to note that accretion varies and the measured luminosity is a snapshot that may not represent the flux illuminating the disk when observed by the SMA. In addition derivations of abundances are needed to confirm that the similarities in flux ratios between the different disks correspond to similarities in abundance profiles. These early results are still indicative of a disconnect between the inner and the outer disk-chemistry evolution.

\section{Discussion} \label{sec:disc}


\noindent In this section the observational results from Paper I and the Southern disks are combined to exploit the total sample size of 12 disks (Table \ref{tbl:det}). Within this sample, CO 2--1 and HCO$^+$ 3--2 emission are detected toward all disks, while the other species are mainly detected toward the lower-luminosity objects. Among the M and K stars, the detections rates are 63\% for DCO$^+$, 63(75)\% for H$_2$CO (both the 4--3 and 3--2 lines), 75(88)\% for N$_2$H$^+$ and 88\% for CN (the strongest 2--1 transition) and HCN. This can be compared with detection rates of 0\% for DCO$^+$ and N$_2$H$^+$, 25\% for H$_2$CO and 50--75\% for CN and HCN toward the F and A stars. Table \ref{tbl:det} shows that the low detections rate toward Herbig Ae stars is not explained by systematic disk size differences, rather it supports the suggestion in Paper I and by \citet{Schreyer08} that many detections depend on the presence of cold midplane regions, which may not be maintained around the more luminous stars in the sample.  Even around A stars, massive enough disks should contain cold midplanes at large radii. The tentative DCO$^+$ detection observed toward the Herbig Ae star AB Aur may be evidence of one such object, though this object does have a residual envelope, which may affect the observed chemistry \citep{Semenov05}. Still it is clearly easier to reach cold enough conditions to activate cold gas phase and grain surface chemical pathways around lower luminosity stars. Colder midplanes toward T Tauri stars compared to Herbig Ae stars do not imply that the upper disk atmospheres are colder as well. In fact the disk atmospheres may even be hotter toward T Tauri stars because of X-ray heating.

It is important to note that this survey only targets simple species and more sensitive surveys are needed to observe the kind of complex chemistry that has been detected toward protostellar regions and in comet comas \citep[e.g.][]{vanDishoeck95,Bottinelli04b,Crovisier04}. This survey suggests that disks around low-mass stars will be chemically richer than disks around more luminous stars in such complex molecules, since many of these species form through an ice chemistry that requires cold disk regions \citep{Garrod08,Oberg09d}. The efficiency at which these ices are released into the gas phase remains to be seen -- a tentative detection of a low H$_2$O column density toward DM Tau suggest that it is lower than predicted by models \citep{Bergin10}. These models, however, do not take into account that H$_2$O ice is probably more difficult to desorb non-thermally compared to organic ices because of layered ice structures, where the H$_2$O ice may be buried underneath the CO, CH$_3$OH, and complex organics ice layer \citep{Oberg10a}.

\subsection{Comparison with previous millimeter studies of disks}

The comparatively high detection rates observed in disks around T Tauri stars compared to Herbig Ae stars were previously hinted at by \citet{Thi04}, \citet{Dutrey07} and \citet{ Henning10} who found HCN, N$_2$H$^+$, and C$_2$H toward T Tauri disks, but not toward Herbig Ae disks in small samples of 3--4 sources. An exception to this trend is the recent discovery of SO toward the Herbig Ae star AB Aur, which has not been detected toward T Tauri stars in comparable observations \citep{Fuente10}. SO is expected to be abundant where grain chemistry is not active and this further supports the observations within the DISCS sample that the chemical richness among the T Tauri stars is associated with different cold chemistry and grain processes. The detection of H$_2$CO toward AB Aur and HD 142527 does however suggest that H$_2$CO formation is not always a grain surface process and it may very well form through different paths in cold and hot disks.

CO and HCO$^+$ emission has been resolved toward a small number of protoplanetary disks previously \citep[e.g.][]{Pietu07,Qi08}. In these studies the HCO$^+$ and CO radial profiles are comparable, at least with respect to their outer radii, which suggests that the different CO 2--1 and HCO$^+$ 3--2 emission profiles toward many of the disks in Table \ref{tbl:size} are mainly due to different excitation conditions of the two lines.

\subsection{The effects of inner disk chemistry drivers on the outer disk}

\noindent With eight M and K stars in the DISCS sample it is possible to explore other drivers of chemical complexity aside from stellar luminosity, which was the main focus of Paper I. 
Accretion rates and disk holes have been observed to affect the inner disk chemistry \citep{Pascucci09,Pontoppidan10} and may {\it a priori} affect the outer disk chemistry as well; holes and gaps may allow more UV radiation to penetrate into the outer parts of the disks and accretion luminosity may affect the radiation driven chemistry in the disk atmosphere (e.g. as traced by CN). The similar line emission ratios toward the large full and transition disks and the classical T Tauri object AS 209 and the weak-line T Tauri object IM Lup (as classified by \citet{Pinte08}) suggest however that neither of these source characteristics affect the physical conditions in the outer disk traced by CN, HCN, DCO$^+$, HCO$^+$, N$_2$H$^+$ and H$_2$CO.

Eight of the DISCS targets are included in infrared molecular surveys probing the inner disk chemistry of HCN and C$_2$H$_2$ \citep{Pascucci09} or H$_2$O, OH, HCN, C$_2$H$_2$ and CO$_2$ \citep{Pontoppidan10}. Herbig Ae systems are less frequently detected in different infrared lines compared to T Tauri systems, but this may be due to strong continuum emission from the more luminous stars \citep{Pontoppidan10}. Apart from the coincidence of both millimeter and infrared line poverty toward luminous stars, there is no positive correlation between inner and outer disk detection rates (Table \ref{tbl:det}). AA Tau and AS 205 are detected in multiple infrared molecular lines, but are the chemically poorest T Tauri systems in millimeter lines. Similarly, the disks rich in millimeter lines, IM Lup, DM Tau, LkCa 15 and GM Aur, are not detected in any of the above infrared molecular lines, except for CO$_2$, which is detected toward IM Lup.  This apparent anti-correlation is however not very informative. The lack of molecules in the inner parts of DM Tau, LkCa 15 and GM Aur may be due to less gas there or to the photodissociation of molecules in optically thin gas \citep{Najita10}. The non-detections in the inner disk of IM Lup suggests that accretion luminosity matters for the inner disk chemistry. This is consistent with the findings of \citet{Pascucci09} for disks around brown dwarfs, which typically have low accretion luminosities and have low detection rates of HCN compared to disks around M and K stars. Finally, the lack of most millimeter transitions toward AA Tau and AS 205 may be due to small emission regions because of inclination (AA Tau) and a small tidally truncated disk (AS 205).

The lack of an observed connection between chemical drivers in the inner disk ($<$15~AU) and the outer disk $>$100~AU makes it difficult to predict what chemical processes that dominate much of the planet-forming zone. Here ALMA will be crucial to investigate the region of influence of the inner disk structure and radiation processes. Herschel H$_2$O, OH and CO observations may also help to improve our understanding of this intermediate region, and it so far seems to indicate that {\it lukewarm} gas-phase H$_2$O and OH are more abundant toward Herbig Ae stars compared to T Tauri stars \citep{Sturm10, Bergin10}. This is consistent with more thermal ice desorption toward more luminous stars and it reinforces the importance of grain surface chemistry to explain chemical variations between different systems.

\subsection{The radiation chemistry dependency on disk radius}

\noindent The radial CN/HCN profile within disks may prove to be a far better radiation tracer compared to the disk averaged emission. The CN and HCN radiation seem to have different radial extents toward several disks. Toward V4046 Sgr and LkCa 15, CN has a 1-2" larger emission area, corresponding to 100s of AU. This is consistent with model predictions of CN and HCN emission, which have CN column densities increasing radially, while HCN abundances either have a flat distribution or disappear completely beyond a certain outer radius \citep{Aikawa02,Jonkheid06}. In \citet{Aikawa02} the HCN abundance does not depend on radius outside of the CO snowline, because HCN abundances are connected to the amount of CO (the dominant reservoir of carbon) in the gas-phase. The CO column is almost constant with radius because it always exists in an intermediate layer that only shifts upwards with radius. In contrast, the CN abundance increases with radius, because the lower density environment in the outer disk favors radical formation. 

This difference between CN and HCN emission regions is enhanced in models that include density tapering in the outer disk. In these models, most CN emission will originate in the outer, very low-density part of the disk, which is completely exposed to the interstellar UV field. CN emission may there originate quite close to the midplane where the densities are still high enough to collisionally excite CN 2-1 ($>10^5$ cm$^{-3}$). Because of the low surface density, there is no protected molecular layer where HCN can exist and HCN emission is confined to the inner disk \citep{Jonkheid06}. The radial CN/HCN profile is therefore a promising tool to distinguish between different radiation fields and disk structures around different types of stars.

The radial CN/HCN profile may be especially useful to determine where in the disk stellar UV regulates the chemistry and where the interstellar radiation field dominates. The rich chemistry around IM Lup shows that neither strong stellar radiation nor a high accretion luminosity is required to produce strong line fluxes. Though chemical models are needed to confirm, this suggests that the interstellar radiation field is more important than stellar UV for the energetically driven chemistry in the outer disk around low-mass stars. This is consistent when comparing with models of the stellar and interstellar contributions to the UV field in the TW Hya disk, which shows that in the disk region resolved by the SMA ($>$300~AU),  the ISRF should dominate \citep{Bergin03}, assuming an A$_{\rm V}\sim0.5$ \citep{Pinte08} and an accretion rate of $<$10$^{-9}$. In higher accretion systems and in disks around more luminous stars the ISRF may never be the main regulator however and this is important to resolve since both gas and dust surface chemical evolution depend on whether the UV field is dominated by lines or a black-body continuum \citep{Bergin03,vanDishoeck06}.

Another potential driver of disk chemistry is X-rays \citep[e.g.][]{Kastner08}. The X-ray flux toward IM Lup is not published and may be a viable alternative to the ISRF for explaining the rich chemistry. Higher radial resolution should here help to resolve whether a central or an external radiation source dominates the chemistry.

\section{Conclusions}

\begin{enumerate}
\item Six protoplanetary disks in the Southern sky (IM Lup, AS 205, AS 209, V4046 Sgr, SAO 206462 and HD 142527) have been surveyed with the SMA for CO, HCO$^+$, DCO$^+$, CN, HCN, DCN, H$_2$CO and N$_2$H$^+$, with a high detection rate toward the four M-K stars and a large chemical variability. N$_2$H$^+$, DCO$^+$, DCN and H$_2$CO are all potential cold chemistry tracers, requiring a massive cold disk midplane to be present at detectable column densities, which explains the high T Tauri detection rates. The A and F stars seem to lack such a cold gas reservoir, at least for long enough timescales
\item In the complete DISCS sample consisting of 8 T Tauri and 4 Herbig Ae objects, CO and HCO$^+$ are detected toward all disks. The detection rates for the other molecules toward the T Tauri disks are: HCN: 88\%, CN: 88\%, DCO$^+$: 63\%, H$_2$CO: 63(75)\%, N$_2$H$^+$: 75(88)\% and DCN: 13\%.  Toward the Herbig Ae objects, HCN is detected in 50\%, CN in 75\% and H$_2$CO in 25\% of the disks, and the other molecules not at all.
\item The disk averaged CN/HCN flux is nearly constant within the complete DISCS sample, despite variations in stellar spectral type between M and A, and orders of magnitude differences in accretion luminosities.
\item The outer disk chemistry is similar for full T Tauri disks and transition disks. There is also no correlation between the inner and outer disk chemistry, nor a visible effect of orders of magnitude differences in accretion rates; the outer disk chemistry seems oblivious to the inner disk.
\item P-V diagrams and moment maps reveal differences in the molecular emission of different species, especially the CN/HCN and DCO$^+$/H$_2$CO ratios, which can be used to model the chemical dependencies on radiation fields and gas versus grain temperatures radially within disks as well as across the sample.
\end{enumerate}

{\it Facilities:} \facility{SMA}

\acknowledgments

\noindent This work has benefitted from discussions with Ewine van Dishoeck and Michiel Hogerheijde and helpful suggestions from the anonymous referee. The SMA is a joint project between the Smithsonian Astrophysical Observatory and the Academia Sinica Institute of Astronomy and Astrophysics and is funded by the Smithsonian Institution and the Academia Sinica. Support for K.~I.~\"O. is provided by NASA through a Hubble Fellowship grant  awarded by the Space Telescope Science Institute, which is operated by the Association of Universities for Research in Astronomy, Inc., for NASA, under contract NAS 5-26555.  C.~E. was supported by the National Science Foundation under Award No.~0901947.

\bibliographystyle{aa}

\newpage

\begin{deluxetable}{lccccccc}
\tabletypesize{\scriptsize}
\tablecaption{Central star and disk data. \label{tbl:star}}
\tablewidth{0pt}
\tablehead{
\colhead{Source} & \colhead{RA} & \colhead{DEC} & \colhead{Sp. type}& \colhead{L$_*$ (L$_{\odot}$)} & \colhead{$\dot{M}$ (10$^{-9}$ M$_{\odot}$ yr$^{-1}$)} &\colhead{PA$_{\rm CO}$ (deg)} 
}
\startdata
IM Lup  	 	&15 56 09.23  	&  $-$37 56 05.9 	&M0 &1.3\tablenotemark{a}&\nodata	&-32--(-42)\tablenotemark{g}	\\
AS 205  	 	&16 11 31.35  	& $-$18 38 25.9  	&K5	&4.0\tablenotemark{b,h}&80\tablenotemark{b,h} &165\tablenotemark{h}	\\
AS 209 	 	&16 49 15.29  	& $-$14 22 08.6  	&K5	&1.5\tablenotemark{c,h}&90\tablenotemark{d,h} &86\tablenotemark{h}	\\
V4046 Sgr	&18 14 10.47  	& $-$32 47 34.5 	&K5	&0.5+0.3\tablenotemark{e}&\nodata	&8\tablenotemark{i}\\
HD 142527	&15 56 41.89  	& $-$42 19 23.3	&F6	&69\tablenotemark{f}	&70\tablenotemark{f} &\nodata	\\
SAO 206462	&15 15 48.95  	& $-$37 08 56.1	&F3	&8\tablenotemark{f}&5\tablenotemark{f} &55\tablenotemark{l}\\
\enddata
\\$^{\rm a}$\citet{Hughes94}, $^{\rm b}$\citet{Prato03}, $^{\rm c}$\citet{Herbig88}, $^{\rm d}$\citet{Johns-Krull00}, $^{\rm e}$\citet{Quast00}, $^{\rm f}$\citet{GarciaLopez06}, $^{\rm g}$\citet{Panic09}, $^{\rm h}$\citet{Andrews09},  $^{\rm i}$\citet{Rodriguez10}, 
$^{\rm k}$\citet{Panic09b}, $^{\rm l}$\citet{Grady09}
\end{deluxetable}

\begin{deluxetable}{lccccccc}
\tabletypesize{\scriptsize}
\tablecaption{\small Spectral setup: 1.4 mm setting (214.670--218.653 GHz,
LSB and 226.671--230.654 GHz, USB)$^{\rm a}$. \label{tbl:setup1}}
\tablewidth{0pt}
\tablehead{
\colhead{Chunk } & \colhead{Frequency range} & \colhead{Channels} & \colhead{Resolution}& \colhead{Lines} & \colhead{Frequency}& \colhead{T$_{\rm up}$}\\
 & (GHz) & & (km s$^{-1}$)& &(GHz)&(K)
}
\startdata
  &                 &      &  LSB              &    &        &  \\
\tableline
S31 & 216.064--216.168 & 256 & 0.56 & DCO$^+$ 3--2 & 216.1126 & 21 \\
S18 & 217.155--217.259 & 256 & 0.56 & DCN 3--2     & 217.2386 & 21 \\
S10 & 217.811--217.915 & 256 & 0.56 & $c$-C$_3$H$_2$ $6_{1\:6}-5_{0\:5}$  & 217.8221 & 39 \\
S06 & 218.140--218.244 & 128  & 1.12 & H$_2$CO $3_{0\:3}-2_{0\:2}$ &
218.2222 & 21\\
S03 & 218.392--218.496 & 256 & 0.56 & CH$_3$OH-E $4_{2\:1}-3_{1\:2}$ &
218.4401 & 46 \\
\tableline
 &                 &      &  USB                &   &        &  \\
\tableline
S04$^{\rm b}$ 	&225.611--225.708	&256&	0.54&H$_2$CO $3_{1\:2}-2_{1\:1}$&225.6978 &33\\
S01 & 226.671--226.775 & 128 & 1.07 & CN $2_{2\:2}-1_{1\:1}$ &
226.6794 & 16 \\
S03 & 226.828--226.932 & 256 & 0.54 & CN $2_{3\:3/4/2}-1_{2\:2/3/1}$ &
226.8747 & 16 \\
S47 & 230.468--230.572 & 256 & 0.53 & CO 2--1 & 230.538 & 17 \\
\enddata
\\$^{\rm a}$ 2 GHz setup for V4046 Sgr (215.413--217.397 GHz, 219.366--221.347 GHz,  225.413--227.397 GHz and 229.366--231.347 GHz) with different chunk assignments.
\\$^{\rm b}$ Only observed toward V4046 Sgr. 
\end{deluxetable}

\begin{deluxetable}{lcccccc}
\tabletypesize{\scriptsize}
\tablecaption{\small Spectral setup: 1.1 mm setting (265.777--269.760 GHz,
LSB and 277.778--281.761 GHz, USB)$^{\rm a}$. \label{tbl:setup2}}
\tablewidth{0pt}
\tablehead{
\colhead{Chunk } & \colhead{Frequency range} & \colhead{Channels} & \colhead{Resolution}& \colhead{Lines} & \colhead{Frequency}& \colhead{T$_{\rm up}$}\\
 & (GHz) & & (km s$^{-1}$)& &(GHz)&(K)
}
\startdata
  &                 &      &  LSB              &    &        &  \\
\tableline
S47 & 265.858--265.962 & 256 & 0.46 & HCN 3--2 & 265.8862 & 26 \\ 
S27 & 267.499--267.603 & 256 & 0.46 & HCO$^+$ 3--2 & 267.5575 & 26 \\
\tableline
 &                 &      &  USB                &   &        &  \\
\tableline
S22 & 279.499--279.603 & 256 & 0.44 & N$_2$H$^+$ 3--2& 279.5117 & 27\\
S46 & 281.500--281.604 & 128& 0.87 & H$_2$CO $4_{1\:4}-3_{1\:3}$ &
281.5269 & 46 \\
\enddata
\\$^{\rm a}$ 2 GHz setup for V4046 Sgr (265.701--267.684 GHz and 275.701--277.684)
\end{deluxetable}

\begin{deluxetable}{rccccccc}
\tabletypesize{\scriptsize}
\tablecaption{\small SMA Observation Log. \label{tbl:obs}}
\tablewidth{0pt}
\tablehead{
\colhead{Date} & \colhead{Sources} & \colhead{Setting\tablenotemark{a}} & \colhead{Baselines\tablenotemark{b} (Antennas)}& \colhead{$\tau_{\rm 225GHz}$} & \colhead{T$_{sys}$(K)\tablenotemark{c}}&\colhead{Beam size\tablenotemark{d} / "} &\colhead{Beam PA\tablenotemark{d} / degrees}
}
\startdata
2009	 Apr 25	&V4046 Sgr	&1.4 mm	&16--139(7)	&0.05--0.1	&156--329 	&2.1$\times$1.4	&3\\ 
	June 22	&V4046 Sgr	&1.4 mm	&6--102(8)	&0.06- 0.11	&78--190		&\nodata 	&\nodata\\
	June 23	&V4046 Sgr	&1.1 mm	&8--125(6)	&0.05--0.1	&121--311	&3.7$\times$2.1	&47 \\
	Sep 9	&V4046 Sgr	&1.1 mm	&5--54(5)		&0.06--0.08	&119--292	&\nodata	&\nodata\\
2010 Apr 26	&SAO 206462	&1.4 mm	&16--77(7)	&0.05--0.1	&78--252		&5.5$\times$3.0	&-5\\
	May 18	&SAO 206462	&1.1 mm	&16--69(7)	&0.06--0.09	&120--295	&5.2$\times$2.5&2 \\
	May 22	&IM Lup		&1.4 mm	&16--77(8)	&0.06--0.08	&102--179	&5.5$\times$3.0	&-1\\
	May 23	&IM Lup		&1.1 mm	&16--69(7)	&0.04--0.06	&124--208	&5.3$\times$2.5	&4\\
	May 25	&HD 142527	&1.4 mm	&16--69(7)	&0.05-0.1		&107--354	&5.5$\times$2.8	&-3\\
	May 26	&HD 142527	&1.1 mm	&16--77(8)	&0.03--0.06	&68--171		&6.6$\times$2.8	&-8\\
	May 31	&AS 209		&1.4 mm	&16--77(6)	&0.05--0.07	&76--163		&3.7$\times$3.6	&-71\\
	June 6	&AS 209		&1.1 mm	&16--69(7)	&0.05-0.11	&98--273		&4.5$\times$2.6	&26\\ 
	June 25	&AS 205		&1.1 mm	&16--77(7)	&0.05--0.09	&114--262	&3.8$\times$2.6	&7\\
	July 30	&AS 205		&1.4 mm	&25--139(6)	&0.05--0.06	&82--175		&3.1$\times$2.0	&53\\
	Aug 4	&V4046 Sgr	&1.1 mm	&16--139(8)	&0.04--0.05	&106--223	&\nodata	&\nodata\\
\enddata
\tablenotetext{a}{See Tables \ref{tbl:setup1} and \ref{tbl:setup2} for 4 GHz settings, 2009 tracks were observed with 2 GHz band width} \tablenotetext{b}{Projected baseline in meters}\tablenotetext{c}{Double sideband (DSB) system temperature.}\tablenotetext{d}{When observed at multiple dates the beam is synthesized from all available data.}
\end{deluxetable}

\begin{deluxetable}{lllllllllllllllllllll}
\tabletypesize{\scriptsize}
\tablecaption{\small Example of IM Lup spectra. Complete spectra for all sources are available on-line. \label{tbl:spec}}
\tablewidth{0pt}
\setlength{\tabcolsep}{0.05in}
\tablehead{
\multicolumn{2}{c}{CO 2-1} & \multicolumn{2}{c}{HCO$^+$ 3-2} & \multicolumn{2}{c}{DCO$^+$ 3-2} & \multicolumn{2}{c}{N$_2$H$^+$ 3-2} & \multicolumn{2}{c}{H$_2$CO 4-3}& \multicolumn{2}{c}{H$_2$CO 3-2}& \multicolumn{2}{c}{HCN 3-2}& \multicolumn{2}{c}{DCN 3-2} & \multicolumn{2}{c}{CN 2$_3$-1$_2$}& \multicolumn{2}{c}{CN 2$_2$-1$_1$}        \\
Vel.& Flux    & Vel.& Flux    & Vel.& Flux    & Vel.& Flux    & Vel.& Flux    & Vel.& Flux    & Vel.& Flux    & Vel.& Flux    & Vel.& Flux    & Vel.& Flux    \\
km/s    & Jy      & km/s    & Jy      & km/s    & Jy      & km/s    & Jy      & km/s    & Jy      & km/s    & Jy      & km/s    & Jy      & km/s    & Jy      & km/s    & Jy      & km/s    & Jy   
}
\startdata
-8.40    &-0.05    &-8.70    &-0.02    &-8.80    &-0.00    &-8.20    &0.04     &-8.50    &-0.11    &-8.80    &0.10     &-8.90    &-0.15    &-8.00    &-0.05    &-8.80    &-0.31    &-8.30    &0.09      \\
-7.90    &-0.17    &-8.30    &0.44     &-8.20    &0.00     &-7.30    &0.09     &-7.60    &0.00     &-8.20    &-0.02    &-8.40    &-0.44    &-6.90    &-0.06    &-8.30    &-0.07    &-7.20    &0.17      \\
-7.40    &0.12     &-7.80    &-0.23    &-7.70    &0.00     &-6.40    &-0.23    &-6.80    &-0.03    &-7.70    &0.04     &-8.00    &0.04     &-5.70    &0.03     &-7.80    &-0.02    &-6.10    &0.01      \\
-6.90    &0.29     &-7.40    &0.19     &-7.10    &0.10     &-5.60    &-0.07    &-5.90    &-0.08    &-7.10    &0.06     &-7.50    &-0.22    &-4.60    &0.10     &-7.20    &-0.09    &-5.10    &-0.15     \\
-6.30    &0.35     &-6.90    &0.09     &-6.60    &0.00     &-4.70    &0.25     &-5.00    &-0.21    &-6.60    &0.12     &-7.10    &0.06     &-3.50    &0.10     &-6.70    &-0.03    &-4.00    &0.02      \\
-5.80    &-0.01    &-6.40    &0.40     &-6.00    &0.07     &-3.80    &0.13     &-4.20    &0.06     &-6.00    &-0.02    &-6.60    &-0.30    &-2.40    &-0.03    &-6.20    &0.08     &-2.90    &-0.07     \\
-5.30    &-0.02    &-6.00    &0.17     &-5.40    &-0.04    &-2.90    &-0.10    &-3.30    &0.02     &-5.40    &-0.19    &-6.10    &-0.08    &-1.30    &0.04     &-5.60    &0.02     &-1.90    &0.07      \\
-4.70    &-0.30    &-5.50    &0.19     &-4.90    &0.14     &-2.10    &-0.01    &-2.40    &-0.16    &-4.90    &0.05     &-5.70    &0.08     &-0.10    &-0.07    &-5.10    &-0.05    &-0.80    &-0.14     \\
\enddata
\end{deluxetable}

\begin{deluxetable}{lcccccc}
\tabletypesize{\scriptsize}
\tablecaption{Continuum flux densities in mJy and spectrally integrated line intensities and 2-$\sigma$ upper limits in Jy km s$^{-1}$ integrated over the disk area. \label{tbl:int}}
\tablewidth{0pt}
\tablehead{
\colhead{Species} & \colhead{IM Lup}&\colhead{AS 205}& \colhead{AS 209}&\colhead{V4046 Sgr}& \colhead{SAO 206462} &\colhead{HD 142527}
}
\startdata
218 GHz			& 199[20]		&345[35]		&349[35]		&359[36]		&135[13]		&2330[230]\\
        \smallskip
267 GHz			&  331[33]		&528[53]		&229[23]		&415[42]		&269[27]		&1030[100]\\
        CO 2-1 		&  20.55[0.20] 	&20.06[0.13] 	&   6.60[0.17] 	&19.98[0.27] 	&  10.39[0.21] 	&  20.76[0.23]\\
   HCO$^+$ 3-2 	&   9.67[0.29] 	& 2.50[0.12] 	&   4.02[0.16] 	&11.43[0.24] 	&   1.18[0.22] 	&   7.79[0.53]\\
   DCO$^+$ 3-2 	&   0.41[0.09]$^{\rm a}$	&$<0.20$ 	&   0.29[0.08]$^{\rm a}$	& 0.80[0.24] 	&  $<0.34$ 	&   $<0.27$\\
N$_2$H$^+$ 3-2 	&   1.59[0.29] 	& $<0.29$ 	&   0.55[0.14]$^{\rm a}$ 	& 2.58[0.20] &  $<0.61$ 	&   $<1.04$\\
   H$_2$CO 4-3 	&   1.37[0.24] 	& $<0.30$ 	&   $<0.48$ 	& 0.95[0.15] &   $<0.44$ 	&   1.97[0.42]$^{\rm a}$ \\
   H$_2$CO 3-2$^{\rm b}$  	&   0.53[0.12] 	& $<0.16$ 	&   $<0.21$ 	& 1.01[0.22] 	&   $<0.30$ 	&   0.27[0.11]$^{\rm a}$ \\
       HCN 3-2 		&   3.51[0.24]	& $<0.34$ 	&   2.01[0.13] 	& 9.89[0.19] 	&   0.58[0.15] 	&   $<0.77$\\
       DCN 3-2 		&  $<0.31$ 	& $<0.18$ 	&   $<0.25$ 	& $<0.38$ 	&   $<0.29$ 	&  $<0.26$\\
CN 2$_3$-1$_2$ 	&   3.55[0.15] 	& $<0.48$	&   3.32[0.14] 	&12.03[0.30] 	&   0.58[0.14] 	&   $<0.26$\\
CN 2$_2$-1$_1$ 	&   $<0.34$ 	& $<0.23$	&   $<0.30$ 	& 1.29[0.33] 	&   $<0.31$ 	&   $<0.34$\\
\enddata
\\$^{\rm a}$Data extracted with a mask covering half the CO gas disk.
\\$^{\rm b}$H$_2$CO $3_{1\:2}-2_{1\:1}$ toward V4046 Sgr, H$_2$CO $3_{0\:3}-2_{0\:2}$ toward all other sources.
\end{deluxetable}

\newpage

\begin{deluxetable}{lcccc}
\tabletypesize{\scriptsize}
\tablecaption{\small FWHM major axes from Gaussian fits in the UV plane of CO, HCO$^+$, HCN and CN line emission with 1-$\sigma$ uncertainties. \label{tbl:size}}
\tablewidth{0pt}
\setlength{\tabcolsep}{0.05in}
\tablehead{
\colhead{Source} &\colhead{CO 2-1} & \colhead{HCO$^+$ 3-2} & \colhead{HCN 3-2} & \colhead{CN 2$_3$-1$_2$}       
}
\startdata
DM Tau	&5.4[0.2]	&3.2[0.3]	&3.3[0.5]	&3.9[0.3]	\\
IM Lup	&5.2[0.2]	&2.8[0.2]	&3.7[0.7]	&3.8[0.5]	\\
AA Tau	&3.4[0.2]	&1.6[0.4]	&3.0[1.3]	&2.5[0.4]	\\
GM Aur	&4.3[0.1]	&2.4[0.3]	&4.7[1.3]	&2.6[0.8]	\\
V4046 Sgr&4.5[0.2]	&3.4[0.6]	&2.0[0.4]	&4.0[0.3]	\\
AS 205	&2.0[0.2]	&1.7[0.6]	&\nodata	&\nodata	\\
AS 209	&2.7[0.8]	&2.4[0.6]	&\nodata	&\nodata	\\
LkCa 15	&3.8[0.2]	&3.2[0.3]	&2.9[0.4]	&4.1[0.4]	\\
HD 142527&4.6[0.2]	&5.4[0.9]	&\nodata	&\nodata	\\
SAO 206462&1.7[0.6]&3.3[1.5]	&\nodata	&\nodata	\\
CQ Tau	&\nodata	&\nodata	&\nodata	&\nodata	\\
MWC 480	&3.3[0.1]	&1.8[0.5]	&3.2[1.1]	&2.1[0.7]	\\
\enddata
\end{deluxetable}

\newpage

\begin{deluxetable}{lllllllllllllllllllll}
\tabletypesize{\scriptsize}
\tablecaption{\small Disk data and line detection rates -- IR non-detections in square brackets \label{tbl:det}}
\tablewidth{0pt}
\setlength{\tabcolsep}{0.05in}
\tablehead{
\colhead{Source}  & \colhead{M$_{\rm D}^{\rm a}$ (M$_{\odot}$)} & \colhead{$d^{\rm b}$ (pc)}&\colhead{r$_{\rm CO}$ (AU)}  & \colhead{detected molecules in DISCS$^{\rm c}$} & \colhead{observed molecules in IR}    
}
\startdata
DM Tau	&0.02	&140	&760	&DCO$^+$ N$_2$H$^+$ H$_2$CO HCN CN	&[HCN C$_2$H$_2$]\\
IM Lup	&0.1		&190	&990	&DCO$^+$ N$_2$H$^+$ H$_2$CO HCN CN	&CO$_2$ [H$_2$O OH HCN C$_2$H$_2$]\\
AA Tau	&0.01	&140	&480	&(N$_2$H$^+$) H$_2$CO HCN CN	&H$_2$O OH HCN C$_2$H$_2$ CO$_2$\\
GM Aur	&0.03	&140	&600	&N$_2$H$^+$ H$_2$CO HCN CN	&[HCN C$_2$H$_2$]\\
V4046 Sgr&0.0033--0.01$^{\rm e}$	&73		&250	&DCO$^+$ N$_2$H$^+$ H$_2$CO HCN CN	&\nodata\\
AS 205	&0.03	&125	&250	&(HCN)	&H$_2$O OH HCN C$_2$H$_2$ CO$_2$\\
AS 209	&0.03	&125	&340	&DCO$^+$ N$_2$H$^+$ HCN CN	&\nodata\\
LkCa 15	&0.05	&140	&530	&DCO$^+$ N$_2$H$^+$ H$_2$CO HCN DCN CN	&[HCN C$_2$H$_2$]\\
HD 142527&\nodata	&150	&690	&H$_2$CO	&[H$_2$O OH HCN C$_2$H$_2$ CO$_2$]\\
SAO 206462&0.01	&84		&140	&HCN CN	&[H$_2$O OH HCN C$_2$H$_2$ CO$_2$]\\
CQ Tau	&0.01	&140	&\nodata	&CN	&\nodata\\
MWC 480	&0.11	&140	&460	&HCN CN	&\nodata\\
\enddata
\\$^{\rm a}$ \citet{Mannings00,Chiang01,Andrews05,Pinte08,Andrews09,Rodriguez10,Grady09}
\\$^{\rm b}$  \citet{Pinte08,Pontoppidan10}
\\$^{\rm c}$ In addition to CO and HCO$^+$
\\$^{\rm d}$ Not including CO \citet{Pascucci09,Pontoppidan10}.
\\$^{\rm e}$ Lower limit based on CO gas and higher limit on dust emission and the assumption of a ISM gas to dust mass ratio \citep{Rodriguez10}.
\end{deluxetable}

\appendix

\section{Position-velocity diagrams for the Taurus disk sample}

\noindent This section contains position-velocity diagrams for the DISCS data presented in Paper I.

\begin{figure}[htp]
\epsscale{1.0}
\plotone{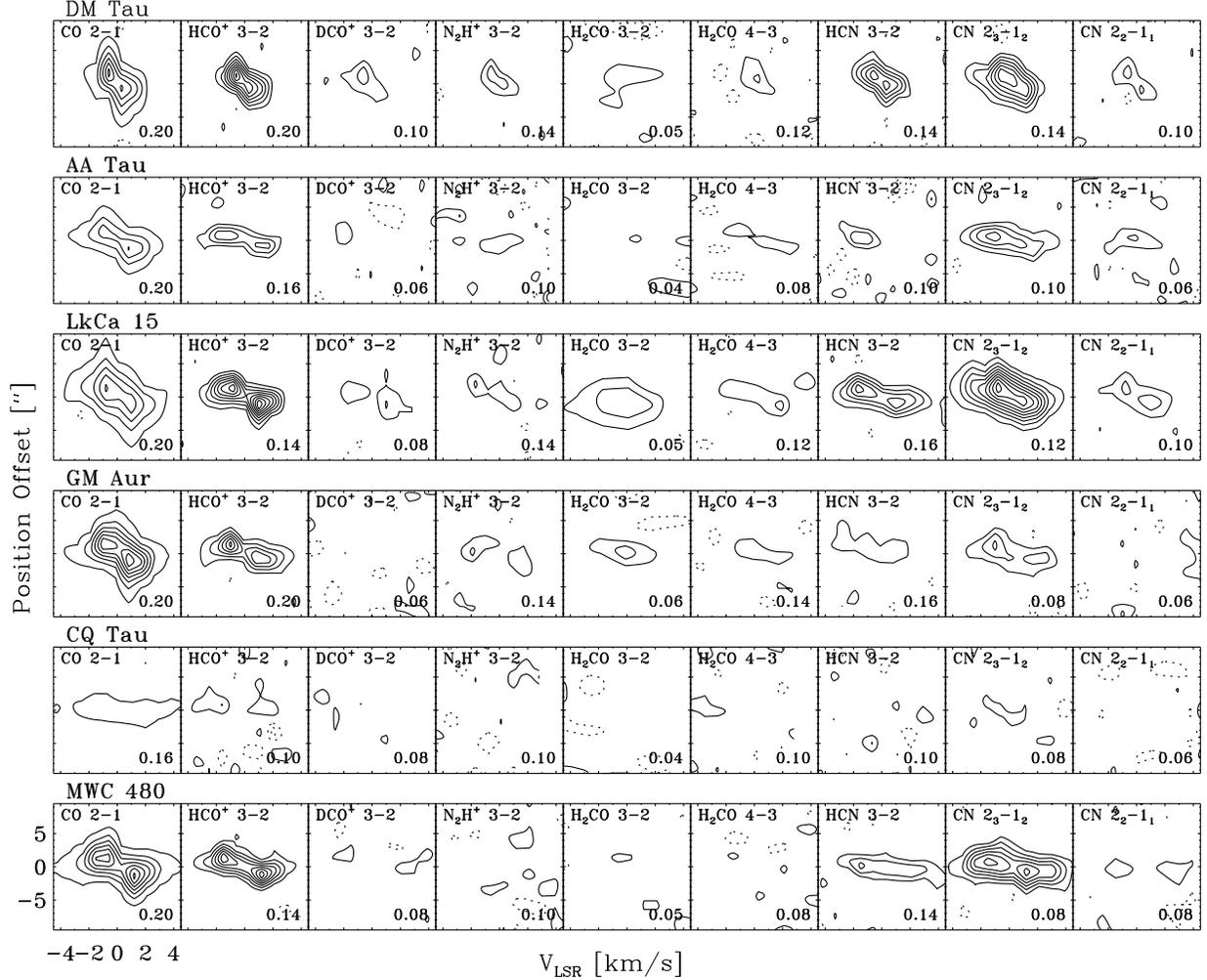}
\caption{Position velocity diagrams toward DM Tau, AA Tau, LkCa 15, GM Aur, CQ Tau and MWC 480. The first contour levels are listed in the bottom right corner of each panel. For all molecules but CO this is also the level of each contour step, while for CO the following steps are three times larger for visibility.  \label{fig:pv_tau}}
\end{figure}

\end{document}